\documentclass[aps,pra,twocolumn,superscriptaddress,floarfix]{revtex4-2}
\usepackage{graphicx}
\usepackage{amssymb}
\usepackage{amsmath}
\usepackage{braket}
\usepackage{appendix}
\usepackage{comment}
\usepackage{cuted}
\usepackage{mathtools}
\usepackage[colorinlistoftodos]{todonotes}
\usepackage[normalem]{ulem}
\usepackage{float}
\usepackage{tikz,siunitx}
\numberwithin{equation}{section}

\usepackage[left]{lineno}
\newcommand{\singletSzero}{{^1\hspace{-.03in}S_0}}

\newcommand{\singletPone}{{^1\hspace{-.03in}P_1}}

\newcommand{\tripletPone}{{^3\hspace{-.03in}P_1}}

\newcommand{\creation}{{\hat{\Psi}^{\dagger}}}
\newcommand{\annihilation}{{\hat{\Psi}}}

\newcommand{\gthree}{{\textsl{g}^{(3)}}}
\newcommand{\gtwo}{{\textsl{g}^{(2)}}}

\newcommand{\gOne}{{G^{(1)}}}
\newcommand{\gTwo}{{G^{(2)}}}
\newcommand{\gThree}{{G^{(3)}}}

\begin{document}

\title{Measuring nonlocal three-body spatial correlations with Rydberg trimers in ultracold quantum gases}

\author{S. K. Kanungo}
\affiliation{Department of Physics and Astronomy, Rice University, Houston, TX  77005-1892, USA}
\affiliation{Rice Center for Quantum Materials, Rice University, Houston, TX  77005-1892, USA}

\author{Y. Lu}
\affiliation{Department of Physics and Astronomy, Rice University, Houston, TX  77005-1892, USA}
\affiliation{Rice Center for Quantum Materials, Rice University, Houston, TX  77005-1892, USA}

\author{F. B. Dunning}
\affiliation{Department of Physics and Astronomy, Rice University, Houston, TX  77005-1892, USA}

\author{S. Yoshida}
\affiliation{Institute for Theoretical Physics, Vienna University of Technology, Vienna A-1040, Austria, EU}
\author{J. Burgd$\ddot{\text{o}}$rfer}
\affiliation{Institute for Theoretical Physics, Vienna University of Technology, Vienna A-1040, Austria, EU}

\author{T. C. Killian}
\affiliation{Department of Physics and Astronomy, Rice University, Houston, TX  77005-1892, USA}
\affiliation{Rice Center for Quantum Materials, Rice University, Houston, TX  77005-1892, USA}

\begin{abstract}

We measure nonlocal third-order spatial correlations in non-degenerate ultracold gases of bosonic ($^{84}$Sr) and spin-polarized fermionic ($^{87}$Sr) strontium through studies of the formation rates for ultralong-range trimer Rydberg molecules. The trimer production rate is observed to be very sensitive to the effects of quantum statistics with a strong enhancement of up to a factor of six (3!) in the case of bosonic $^{84}$Sr due to bunching, and a marked reduction for spin-polarized fermionic $^{87}$Sr due to anti-bunching. The experimental results are compared to theoretical predictions and good agreement is observed. The present approach opens the way to {\it{in situ}} studies of higher-order nonlocal spatial correlations in a wide array of ultracold atomic-gas systems. 
\end{abstract}
\maketitle
\section{INTRODUCTION}

Measurements of atom-atom spatial correlations have played an important role in understanding the properties of quantum many-body systems and their non-classical behaviors, such as Bose-Einstein condensates\cite{burt1997}, the Mott insulator state\cite{carcy2019}, quantum spin models and magnetism\cite{browaeys2020,semeghini2021,mazurenko2017,hart2015}, Efimov physics\cite{fletcher2017}, and strongly interacting gases in one dimension \cite{tolra2004,kinoshita2005,haller2011}.
Short-range or local two- and three-body correlations in quantum gases have been examined on length scales $\leq$ 20 nm through studies of photoassociation and three-body recombination\cite{burt1997,tolra2004,kinoshita2005,haller2011}. Long-range or nonlocal correlations with length scales on the order of the wavelength of light have been explored using Bragg spectroscopy\cite{hart2015} and direct imaging in optical tweezers \cite{browaeys2020} and quantum gas microscopes\cite{mazurenko2017,bakr2009}. The measurement of two-body correlations at intermediate length scales, $\sim 20-200$ nm, has been  achieved recently through studies of the formation of ultralong-range Rydberg molecules (ULRMs)\cite{whalen2019}. While this earlier work focussed on the creation of dimer molecules and two-body correlations, we demonstrate here that this approach can be extended to examine higher-order correlations and report measurements of nonlocal three-body spatial correlations in ultracold gases of bosons ($^{84}$Sr) and spin-polarized fermions ($^{87}$Sr).

\begin{figure}[ht]
    \centering
    \includegraphics[width = 0.45\textwidth]{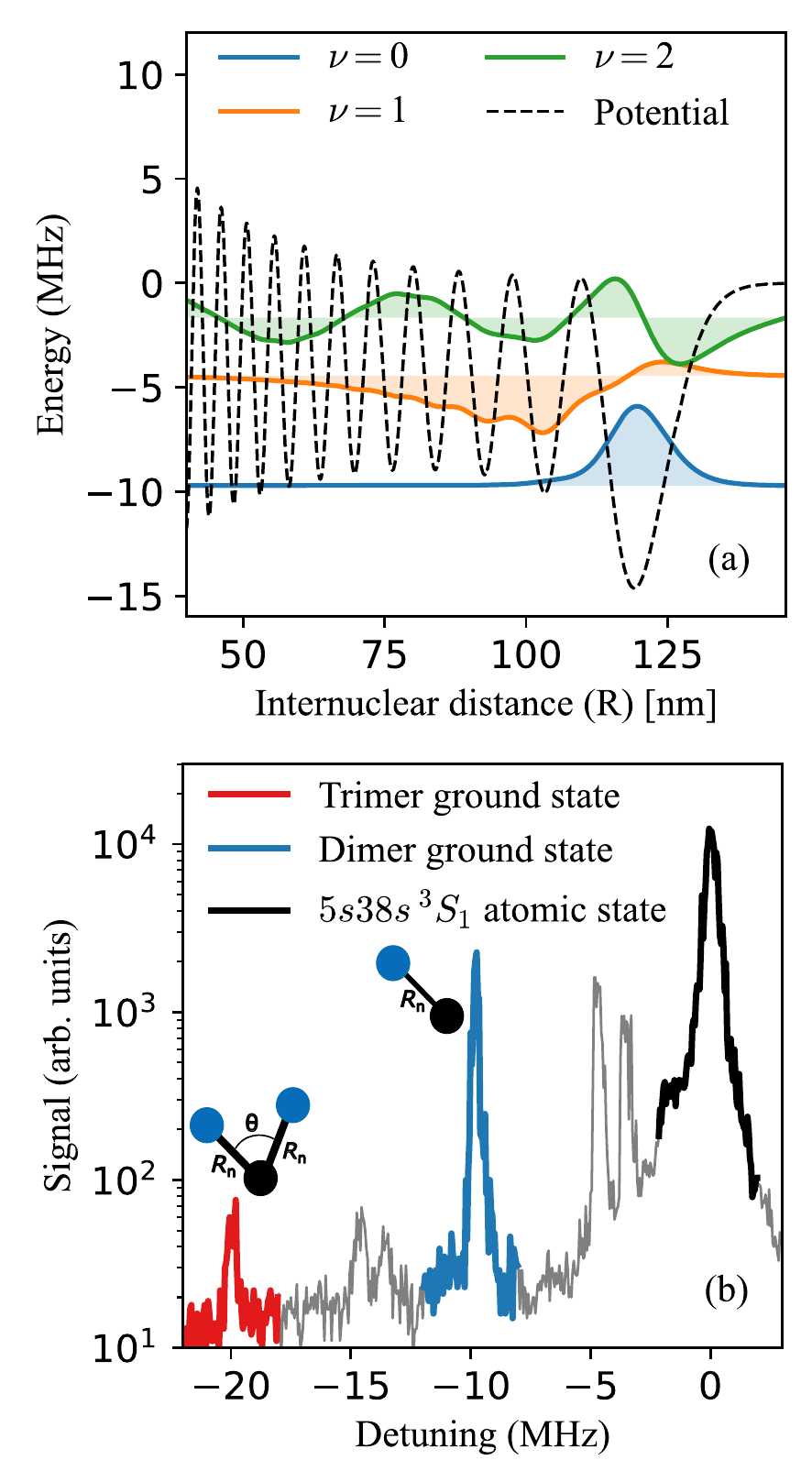}
    \caption{(a) Calculated molecular potential for a $5s38s\,^3S_1-5s^2\,^1S_0$ strontium atom pair. The calculated vibrational wavefunctions, multiplied by the radial coordinate $R$, for the $\nu=$ 0, 1 and 2 vibrational states are included and the horizontal axis for each indicates its binding energy. (b) Rydberg excitation spectrum in a cold dense strontium gas in the vicinity of the $5s38s\,^3S_1$ Rydberg state, where, the $x$-axis shows the laser detuning from the Rydberg atomic line (black). The features associated with the formation of dimer and trimer ground state molecules are shown in blue and red respectively. The other remaining features (gray) correspond to creation of vibrationally-excited molecular states. Illustrations of dimer and trimer molecules accompany the states of interest.}
    \label{fig:Fig1}
\end{figure}

ULRMs are formed through scattering of the Rydberg electron from a ground-state atom embedded within the electron cloud, which results in an attractive ``molecular" potential\cite{greene2000, bendkowsky2009}. A typical example of such a potential for a strontium $5s38s\,^3S_1-5s^2\,^1S_0$ atom pair, calculated using a Fermi pseudopotential, is shown in Fig.\ \ref{fig:Fig1}(a) and mirrors the radial electron probability density distribution. This potential can support a number of vibrational levels, and the vibrational wavefunctions associated with the lower levels are included in Fig.\ \ref{fig:Fig1}(a). Of particular interest here is the ground $\nu=0$ vibrational state, which is strongly localized in the outermost potential well at an ($n-$dependent) internuclear separation $R_n\sim1.8(n-\delta)^2a_0$, where $a_0$ is the bohr radius and $\delta$ is the $s$-state quantum defect. 

Measurements of two-body correlations using dimer formation rates \cite{whalen2019} exploited the fact that the likelihood of creating a dimer in the ground  vibrational state is proportional to the probability that there are two ground-state atoms in the initial sample with the appropriate initial separation, $R_n$. Thus, by varying $n$, and hence $R_n$, it is possible to probe the probability distribution of atomic separations and derive the pair-correlation function $\gtwo(r)$. A trimer ULRM in its ground state contains two ground-state atoms in the vibrational ground state at a distance $R_n$ from the Rydberg core ion.
Measurements of trimer formation can therefore be used to examine three-body spatial correlations, although analysis of the data is more complex than for dimers because the relative positions of the two bound ground state atoms, characterized by the angle $\theta$ shown in Fig. 1(b), is not fixed and the measured values represent angle-averaged quantities. Here results are presented for ULRMs with values of $n$ in the range $29-45$, which correspond to values of $R_n$ of $\sim60-170$ nm, and sample temperatures of 200 nK to 2 $\mu$K. The length scales probed are less than or on the order of the atomic thermal de Broglie wavelength, $\lambda_{\text{dB}}$, $R_n/\lambda_{\text{dB}}\sim0.2-1$, where the effects of quantum statistics should be clearly visible in the correlation functions. 

\section{EXPERIMENTAL METHODS}
In the present work, Rydberg trimer excitation rates are measured in ultracold gases. Cold samples of $^{84}$Sr (boson, nuclear spin $I=0$) and spin-polarized $^{87}$Sr (fermion, $I=9/2$) are prepared using standard methods of laser cooling and trapping\cite{stellmer2013,deescobar2009}. Atoms in an atomic beam are slowed in a Zeeman slower and cooled to $\sim$1 mK in a magneto-optical trap (MOT) using the $5s^2\, \singletSzero \longrightarrow 5s5p\, \singletPone$ transition at 461 nm. To further reduce the temperature to a few $\mu$K, a MOT operating on the narrow $5s^2\, \singletSzero \longrightarrow 5s5p\, \tripletPone$ transition at 689 nm is employed. The atoms are then loaded into an optical dipole trap (ODT) formed by two crossed 1064 nm laser beams, and evaporative cooling\cite{ohara2001} is used to create samples with final temperatures in the range of 200 nK - 2 $\mu$K (evaporative cooling of spin-polarized $^{87}$Sr is performed with $^{84}$Sr present in the trap to provide sympathetic cooling). 
Typically for both isotopes, 2-7 $\times 10^5$ atoms remain trapped in the ODT with peak densities in the range of 0.5 - 2.5 $\times 10^{12}$ cm$^{-3}$ calculated from the measured trap oscillation frequencies and atom number.

To obtain cold samples of spin-polarized fermions, the ground state $^{87}$Sr atoms are optically pumped to the $m_F$ = 9/2  ($F=9/2$) state\cite{whalen2019}. A 7.6\,G bias magnetic field is applied after loading atoms into the ODT, which produces a Zeeman splitting of $\sim$ 650 kHz between adjacent magnetic sublevels in the $5s5p\,^3P_1\,F=9/2$ manifold. Population is transferred to the $m_F = 9/2$ ground state by applying a series of $\sigma^{(+)}$ polarized 689 nm laser pulses that are red-detuned from each of the $m_F\longrightarrow m_F+1 $ transitions by 50 kHz. The resulting spin-polarized $^{87}$Sr atom sample is then sympathetically cooled with $^{84}$Sr atoms to the desired final temperature and the magnetic field is lowered to 1\,G to maintain a quantization axis during subsequent measurements. All data for the $^{84}$Sr and unpolarized $^{87}$Sr atom samples are recorded in zero magnetic field. 

Strontium Rydberg dimers and trimers are created by two-photon excitation from the ground state via the $5s5p\, \tripletPone$ ($F=11/2$ for $^{87}$Sr)  intermediate state. The first (689 nm) photon is blue-detuned 14 MHz from the intermediate state. The second (320 nm) photon is scanned to generate molecular excitation spectra. Both the lasers are switched on for $\sim$10 $\mu$s. The ground-state trimer excitation is spectroscopically resolved from excitation to all other states. Less than one Rydberg molecule is created per laser shot to avoid Rydberg-Rydberg interactions. The Rydberg molecule is detected by selective field ionization\cite{gallagher1994}, and the product electrons are directed to a micro-channel plate (MCP) for detection.
ULRM states with Rydberg principal quantum number $29\le n \le 45$ are created in this study, corresponding to Rydberg atom and ULRM sizes $63\, {\rm{nm}} \le R_n \le 165\, {\rm{nm}}$. Larger $n$ and $R_n$ are not currently accessible because ULRM spectral features become unresolved at the current spectral resolution of $100\,$kHz.
Typically, 1000 experimental cycles can be performed using a single ultracold sample to build up statistics.

\section{$\gthree(r)$ and $\gtwo(r)$ correlation functions}
Measurement of correlation functions provide an effective means 
to examine the behavior of complex quantum systems, in particular
many-body systems. $G^{(p)}({\bf r}_1, \dots,{\bf r}_p)$ $({p} \ge 2)$
represents the diagonal elements of the reduced $p$-body density matrix and measures the likelihood of finding $p$ particles at the specified position at a given time.  The reduced one-particle density matrix (RDM) $\rho^{(1)}(\bf{{r}_1},\bf{{r}_2})$, sometimes also denoted by $G^{(1)}({\bf r}_1, {\bf r}_2)$ \cite{naraschewski1999}, contains information on coherences and off-diagonal correlations which contrasts the $G^{(p)}$ for values of $p \ge 2$. The theoretical description of
correlation functions in the atomic physics context was explored by Glauber et al. [19] and we follow that analysis  to derive the angle-averaged three-body correlation function relevant for trimer ULRM excitation. In the following we will show that for an ideal gas the ensemble
averaged three-body correlation function can be expressed solely in terms of the RDM.
    
Let $\hat{\Psi}^{\dagger}(\mathbf{r})$ and $\hat{\Psi}(\mathbf{r})$ be the creation and annihilation operators for an atom at position ${\mathbf{r}}$, which obey commutation relations appropriate to either bosons or fermions. The RDM is then given by:
\begin{equation}
\label{eq:G1}
G^{(1)}(\mathbf{r_1,r_2}) = \braket{\creation(\mathbf{r})\annihilation(\mathbf{r_2})},
\end{equation}
and its diagonal elements represent the density:
\begin{equation}
\rho (\mathbf{r}) = G^{(1)}(\mathbf{r,r}).
\end{equation}
$G^{(1)}(\mathbf{r_1,r_2})$ for atoms trapped in a potential $V(\mathbf{r})$ can be expressed in terms of the generalized Bose function, $g_{\alpha}$\cite{naraschewski1999}, as
\begin{widetext}
\begin{equation}\label{Firstordercorrexpr}
    \gOne(\mathbf{r_1,r_2})=
    \frac{1}{\lambda_{\text{dB}}^3}g_{3/2}\left(\text{exp}\left[ \frac{\mu-[V(\mathbf{r_1})+V(\mathbf{r_2})]/2}{k_BT}\right],\text{exp}\left[-\pi\frac{(\mathbf{r_2-r_1})^2}{\lambda^2_{\text{dB}}} \right] \right),
\end{equation}
\end{widetext}
where $g_{\alpha}$ is given by the series,
\begin{equation}
g_\alpha(x,y) = \sum_{k=1}^{\infty}\frac{x^ky^{1/k}}{k^\alpha},
\end{equation}
$\mu$ is the chemical potential, and $\lambda_{\text{dB}}$ is the thermal de-Broglie wavelength determined by the sample temperature $T$.
The optical dipole trap in the present experiment is  well approximated by the anisotropic harmonic potential,
\begin{equation}\label{harmonicpot}
    V(\mathbf{r}) = \frac{m(\omega_x^2x^2+\omega_y^2y^2+\omega_z^2z^2)}{2},
\end{equation}
where $m$ is mass of the strontium atom, and $\omega_x$, $\omega_y$ and $\omega_z$ are the trap oscillation frequencies. The sizes of the Rydberg molecules studied here are small compared to the trap dimensions and it is reasonable to assume that $V(\mathbf{r_1})  \approx V(\mathbf{r_2}) \approx V(\mathbf{r})$.
Use of Eq. (\ref{harmonicpot}) in Eq. (\ref{Firstordercorrexpr}), allows numerical calculation of $\gOne(\mathbf{r_1,r_2})$.

Expressions for $\gTwo(\mathbf{r_1,r_2})$ and $\gThree(\mathbf{r_1,r_2,r_3})$ can be written as
\begin{equation}
\label{eq:G2}
G^{(2)}(\mathbf{r_1,r_2}) = \braket{\creation(\mathbf{r_1})\creation(\mathbf{r_2})\annihilation(\mathbf{r_2})\annihilation(\mathbf{r_1})},
\end{equation}
\begin{equation}
\label{eq:G3}
G^{(3)}(\mathbf{r_1,r_2,r_3}) =
\braket{\creation(\mathbf{r_1})\creation(\mathbf{r_2})\creation(\mathbf{r_3})\annihilation(\mathbf{r_3})\annihilation(\mathbf{r_2})\annihilation(\mathbf{r_1})},
\end{equation}
where $\mathbf{r_1,r_2}$ and $\mathbf{r_3}$ denote the position vectors of the various particles. For an ideal gas, using Wick's theorem\cite{wick1950}, $\gThree(\mathbf{r_1,r_2,r_3})$ can be expressed in terms of one-body density matrices\cite{hodgman2011} as
\begin{equation}
    \begin{aligned}
    G^{(3)}(\mathbf{r_1}&,\mathbf{r_2},\mathbf{r_3}) = \gOne(\mathbf{r_1,r_1})\gOne(\mathbf{r_2,r_2})\gOne(\mathbf{r_3,r_3}) \\
    &\pm |\gOne(\mathbf{r_1,r_2})|^2\gOne(\mathbf{r_3,r_3}) \\
    &\pm |\gOne(\mathbf{r_2,r_3})|^2\gOne(\mathbf{r_1,r_1}) \\
    &\pm |\gOne(\mathbf{r_3,r_1})|^2\gOne(\mathbf{r_2,r_2})\\
    &+2\text{Re}\left\{(\gOne(\mathbf{r_1,r_2})\gOne(\mathbf{r_2,r_3})\gOne(\mathbf{r_3,r_1})\right\}.
    \end{aligned}
\end{equation}
The corresponding expression for $\gTwo(\mathbf{r_1,r_2})$ is:
\begin{equation}
    G^{(2)}(\mathbf{r_1,r_2}) = \gOne(\mathbf{r_1,r_1})\gOne(\mathbf{r_2,r_2}) \pm |\gOne(\mathbf{r_1,r_2})|^2.
\end{equation}
In the above expressions, the + (-) sign applies to identical bosons (fermions) in the same internal state.

Consider a dimer molecule where the position vectors $\mathbf{r_1}$ and $\mathbf{r_2}$ correspond to the atoms that comprise the Rydberg-bound ground-state atom pair of the dimer, and $\mathbf{r} = \mathbf{r_2} - \mathbf{r_1}$ denotes the position of the  ground-state atom relative to the core ion. $\mathbf{R} = (\mathbf{r_2} + \mathbf{r_1})/2$  denotes the position of the center of mass (COM) of the pair. 
$\gTwo$ may then be expressed in terms of these new variables. The normalized and trap-averaged pair correlation function is then given by

\begin{equation}
    \gtwo(r) = \frac{\int d\mathbf{R}\,\gTwo(\mathbf{R},r)}{\int d\mathbf{R}\, \rho(\mathbf{R})^2},
    \label{g2}
\end{equation}
where we have exploited the fact that in the length scale $r$ probed by the dimer,
$\gTwo$ can be approximated as independent of the orientation of {\bf{r}} because  $R_n$ is much smaller than the scale of variation of the trapping potential.

It is less straightforward to define such averaged corelation functions when three bodies are involved, as is the case for a Rydberg trimer. In a ground state Rydberg trimer both of the two ground-state atoms are bound at the same inter-nuclear distance ${r}\approx R_n$ from the core ion. The measured $\gThree$ depends on the angle between the relative orientations of ground-state atoms, defined as the polar angle $\theta$ in Fig. 1, but not on the absolute orientation of the molecule nor on the azimuthal angle.   We consider the relevant three-body correlation function for the experiments described here, which is normalized, trap-averaged, and averaged over $\theta$,
\begin{equation}
    \gthree(r) = \frac{\int d\mathbf{R}\,\left\langle\gThree(\mathbf{R},r,\theta)\right\rangle_{\theta}}{\int d\mathbf{R}\, \rho(\mathbf{R})^3}.
   \label{g3}
\end{equation}

\begin{figure}[hb]
    \centering
    \includegraphics[width=0.46\textwidth]{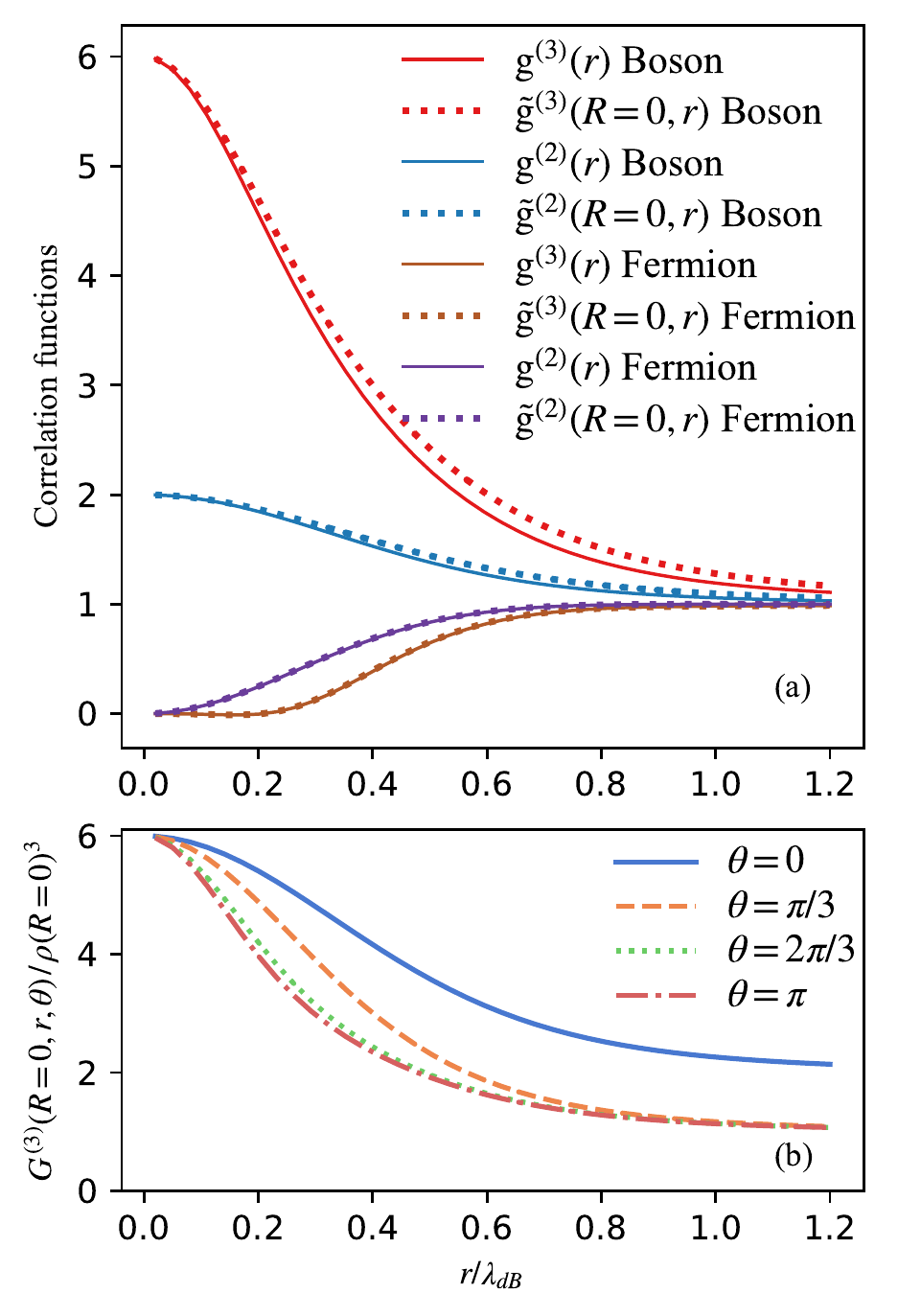}
    \caption{(a) Numerically calculated $\gthree(r)$  and $\gtwo(r)$
    and
    $\tilde{\textsl{g}}^{(3)}(\bold{R},r)
    \equiv{\left\langle\gThree(\mathbf{R},r,\theta)\right\rangle_{\theta}}/{ \rho(\mathbf{R})^3}$ 
    and $\tilde{\textsl{g}}^{(2)}(\bold{R},r)
    \equiv{\gTwo(\mathbf{R},r)}/{ \rho({\bf{R}})^2}$ 
   evaluated  at the trap center $(\bold{R}=0)$,
     for Bose and Fermi gases. (b) Plot of  $\gThree(\mathbf{R},r,\theta)/ \rho(\mathbf{R})^3$ at $\mathbf{R}=0$ as a function of $r/\lambda_{\text{dB}}$ for an ideal gas of bosons
    at various values of $\theta$. At $\theta=0$, bunching in bosons (or anti-bunching in fermions) is expected. In particular, at large $r$, the third atom becomes uncorrelated from the other two and the $\theta-$dependence of the three-body correlation represents the two-body correlation.}
    \label{fig:g2andg3plot}
\end{figure}

Equations (\ref{g2}) and (\ref{g3}) can be numerically evaluated and the results for a trap containing a 500 nK sample at a fugacity of 0.99 are shown in Fig. \ref{fig:g2andg3plot}(a). Fifty terms are retained in the expansion for
$g_\alpha(x,y)$, which is sufficient for convergence of Eq. (\ref{Firstordercorrexpr}) even for a fugacity this close to degeneracy (i.e., 1). As shown in Fig 2, the values of 
${\left\langle\gThree(\mathbf{R},r,\theta)\right\rangle_{\theta}}/{ \rho(\mathbf{R})^3}$ 
and ${\gTwo(\mathbf{R},r)}/{ \rho({\bf{R}})^2}$ 
calculated for the trap center, i.e. $\bold{R}=0$, do not deviate significantly from the trap-volume-averaged values for the present experimental trap parameters. 

Fig.\ \ref{fig:g2andg3plot}(b) shows 
$\gThree(\mathbf{R}=0,r,\theta)/ \rho(\mathbf{R}=0)^3$
for representative values of $\theta$, for an ideal gas evaluated at the coordinates corresponding to a Rydberg trimer. It is interesting to note that the three-body correlation function approaches 2 at large $r$ as $\theta\longrightarrow 0$ and the two atoms that will become the ground-state atoms bound to the  Rydberg core come closer to each other. 

\section{RESULTS AND DISCUSSION}
Generalizing the formalism of \cite{whalen2019}, the measured dimer ($\mathcal{S}^{(\text{2})}_{n}$) and trimer ($\mathcal{S}^{(\text{3})}_{n}$) ULRM signals for principal quantum number $n$,  which we take as the integrals of the photoexcitation spectral lines, depend on several experimental parameters and  may be approximated as:
\begin{equation}
\label{eq: signalrate2}
\begin{aligned}
    \mathcal{S}^{(\text{2})}_{n} \simeq \alpha I_1I_2\beta_{{n}}\mathcal{CO}_{n}^{(2)} \int d^3\mathbf{R}\, 
   \gTwo(\mathbf{R},R_n) \\
   = \alpha I_1I_2\mathcal{N}^{(2)}\beta_{n}\mathcal{CO}_{n}^{(2)}\textsl{g}^{(2)}(R_n).
\end{aligned}
\end{equation}
\begin{equation}
\label{eq: signalrate3}
\begin{aligned}
    \mathcal{S}^{(\text{3})}_n \simeq \alpha I_1I_2\beta_{n}\mathcal{CO}_{n}^{(3)} \int d^3\mathbf{R}\, 
   \left\langle\gThree(\mathbf{R},R_{n},\theta)\right\rangle_{\theta} \\
   = \alpha I_1I_2\mathcal{N}^{(\text{3})}\beta_{n}\mathcal{CO}_{n}^{(3)}\textsl{g}^{(3)}(R_n).
\end{aligned}
\end{equation}
The MCP detection efficiency is characterized by $\alpha$ and is independent of isotopes, and $I_1 \text{ and } I_2$ are the intensities of the Rydberg excitation lasers, which are monitored by photo-detectors. The local excitation rate is assumed to be proportional to $G^{({p})}$, with appropriate arguments and angle average. This quantity is proportional to 
the $p$-th power of the local cold atom density. The integral over the trap results in the nonlocal spatial correlation function, ${\textsl{g}}^{({p})}{(R_n)}$, and the density scaling factor  $\mathcal{N}^{({p})} = \int d^3\mathbf{R}\,\rho(\mathbf{R})^{{p}}$.
Other factors that influence the photoexcitation rate are the square of the reduced two-photon electronic-transition matrix element, represented by $\beta_n$, which depends on the principal quantum number $n$, the ($n-$independent) Clebsch-Gordan coefficients, $\mathcal{C}$, that couple the levels of interest, and the effective Franck-Condon factor, $\mathcal{O}_n^{({p})}$, given by the overlap of the initial scattering wavefunction with the molecular bound state, which is different for dimers and trimers and also depends on $n$\cite{whalen2019}. Once these factors are taken into account any remaining variations in the excitation rates can be attributed to changes in  ${\textsl{g}}^{({p})}{(R_n)}$.
 
\begin{figure}[h]
    \centering
    \includegraphics[width = 0.45\textwidth]{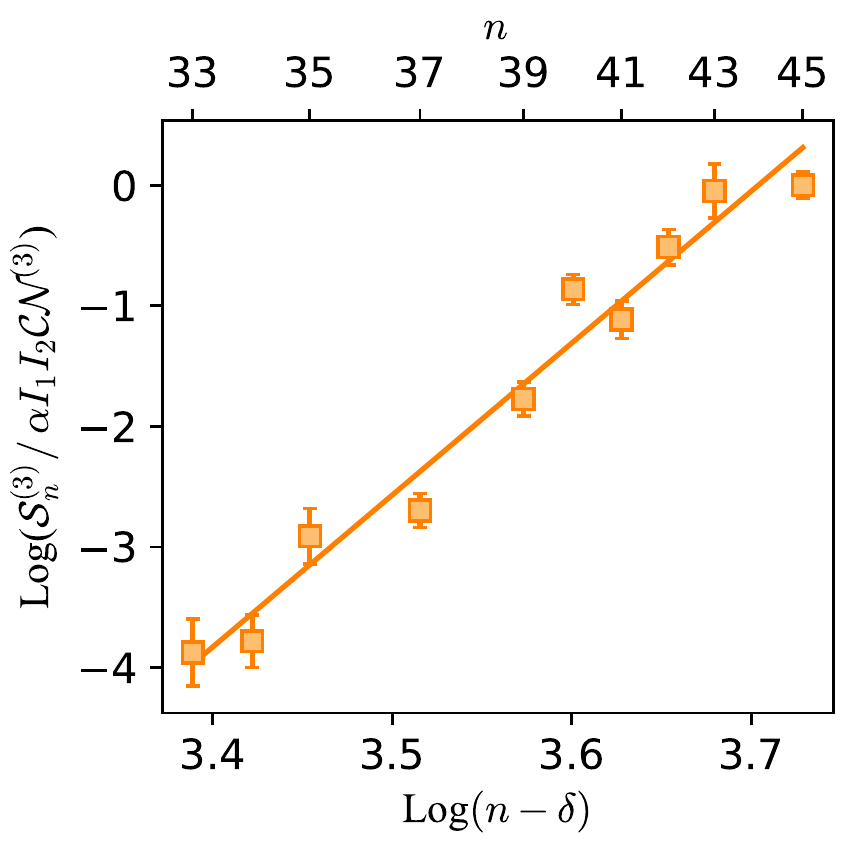}
    \caption{Normalized $n$-dependence of the trimer ground state production rate ($\mathcal{S}^{(\text{3})}_n/ \alpha I_1I_2\mathcal{C} \mathcal{N}^{(\text{3})}$) in an unpolarized Fermi gas of $^{87}$Sr. The measured trimer signals are well fit by a power law with exponent 12.3 $\pm$ 0.8, which furnishes the scaling of the product $\beta_n \mathcal{O}_n^{(\text{3})}$ for trimer excitation. Effects of residual three-body correlations in the  unpolarized sample of $^{87}$Sr are small ($\textsl{g}^{(3)}_{\text{unpol}}(r)\approx$1) and are neglected in determining the normalized trimer excitation rates.}
    \label{fig:trimerFCscaling}
\end{figure}

The $n$-dependence of the product  $\beta_n \mathcal{O}_n^{({p})}$ for trimers and dimers is experimentally determined by measuring molecular excitation rates in an unpolarized $^{87}$Sr sample (see Fig. \ref{fig:trimerFCscaling}). $^{87}$Sr has ten degenerate ground states, and an unpolarized sample approximates a classical gas. 
Ancillary calculations suggest that residual two- and three-body correlations are indeed small, i.e., ${\textsl{g}}^{(p)}_{\text{unpol}}{(R)}\approx 0.9-1$, and their effects are therefore neglected in the calculation of the $n$-dependence of  $\beta_n \mathcal{O}_n^{({p})}$.
For unpolarized $^{87}$Sr, all the factors in Eq.\ \ref{eq: signalrate3} that influence the trimer production rate except $\beta_n$ and $\mathcal{O}_n^{(\text{3})}$ can be measured and taken into account, and thus any $n-$dependence seen in the trimer production rate must be associated with the product $\beta_n \mathcal{O}_n^{(\text{3})}$. In earlier measurements of ground-state dimer production, the product $\beta_n \mathcal{O}_n^{(\text{2})}$ was observed to scale as $(n-\delta)^{3.5(3)}$\cite{whalen2019}, a result confirmed in the present work. As illustrated by Fig 3, measurements of trimer formation showed a much stronger $n-$dependence in the product $\beta_n \mathcal{O}_n^{(\text{3})}$, which scales as $(n-\delta)^{12.3(8)}$.

Figures 4(a-h) illustrate the $n-$dependence of the dimer and trimer excitation spectra recorded using $^{84}\text{Sr}$ and spin-polarized $^{87}\text{Sr}$. In each set of measurements the results are normalized by laser intensities and ground-state atom densities as well as the $n-$dependence in the product $\beta_n \mathcal{O}_n^{({p})}$. Thus any changes seen in the measured signal levels must be associated with changes in the correlation functions ${\textsl{g}}^{(2)}{(R_n)}$ or ${\textsl{g}}^{(3)}{(R_n)}$.

For $^{84}\text{Sr}$ the (normalized) trimer photoexcitation rate increases dramatically with decreasing $n$, pointing to a similar increase in ${\textsl{g}}^{(3)}{(r)}$. This results because, as $n$ decreases, molecule formation probes correlations on ever shorter length scales, which for the present sample temperatures become smaller than the atomic de Broglie wavelength. In this regime bunching in bosons results in an increase in the spatial correlation. As seen in Fig. 4, the dimer production rate for $^{84}\text{Sr}$ also increases as $n$ decreases, but this increase is much less pronounced than that observed for trimers highlighting  how much more sensitive trimer formation is to the effects  spatial correlations. In contrast, the trimer excitation rate in spin-polarized $^{87}\text{Sr}$ decreases significantly with decreasing $n$ due to anti-bunching, i.e., Pauli exclusion. This decrease is more pronounced than that seen in dimer production, further demonstrating the greater sensitivity of trimer production to spatial correlations.

\begin{figure}[H]
    \centering
    \includegraphics[width = 0.45\textwidth]{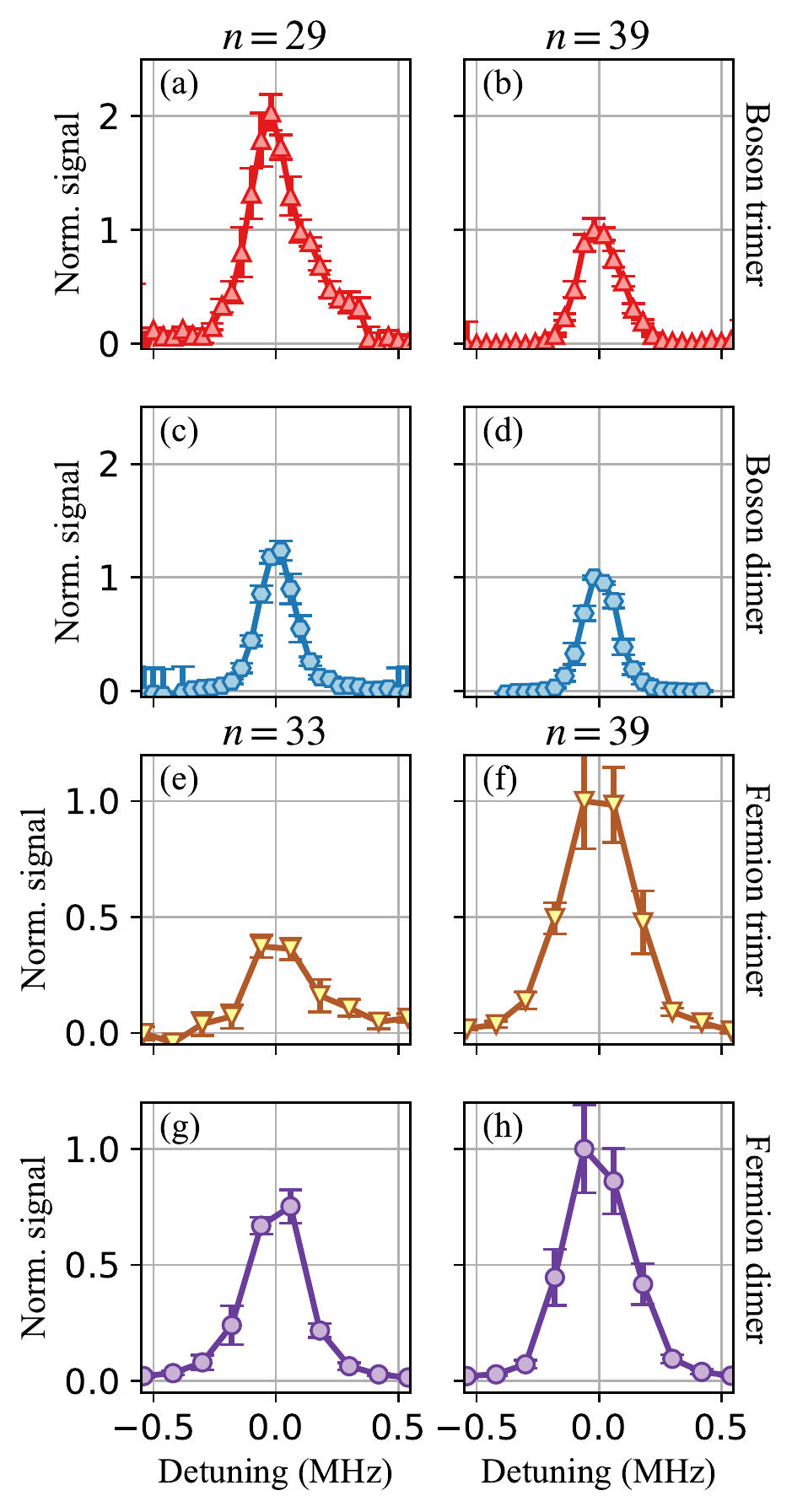}
    \caption{
    Photoexcitation spectra for ground-state dimer and trimer molecules in a cold strontium gas as a function of laser detuning from the dimer or trimer line center. Each data set is normalized by laser intensities, atom densities, and the product $\beta_n \mathcal{O}_n^{({p})}$ (see text). Trimer photoexcitation rates for bosonic $^{84}$Sr are shown in (a,b) and for spin-polarized $^{87}$Sr in (e, f).  Note the strong signal enhancement as $n$ decreases for bosons compared to suppression for fermions. For comparison, measurements of dimer formation in the same gases are plotted in (c, d) and (g, h), which show similar but weaker variation. The data for dimer and trimer production at $n$=39 are all normalized to the same peak height. Note the change in scale of the vertical axes between the upper and lower data sets.
    }
    \label{fig:gallery}
\end{figure}
Figure 5 shows the principal findings of this paper. Dimer and trimer ground state photoexcitation spectra were recorded for a range of quantum numbers, $29\leq n\leq45$, at various temperatures. For each spectrum, the total integrated molecular signal ($\mathcal{S}^{({p})}_n$)
was obtained by fitting  to a voigt profile. The signal was normalized by $\alpha I_1I_2\mathcal{C} \mathcal{N}^{({p})}\beta_n \mathcal{O}_n^{({p})}$
to remove all dependences other than
${\textsl{g}}^{(3)}{(r)}$ or ${\textsl{g}}^{(2)}{(r)}$ [Eqs. (\ref{eq: signalrate2})-(\ref{eq: signalrate3})], where $r = R_n$ is taken to be the size of the ULRM. 
 To enable direct comparison between measurements undertaken at different sample temperatures, the length  is scaled by the atomic thermal de Broglie wavelength, $\lambda_{\text{dB}}$. A single amplitude scaling parameter is fit for each experimental data set, which normalizes the data to match the corresponding theoretical curves calculated from Eqs. (\ref{g2}) and (\ref{g3}). 
 ({{Fermion dimer data is taken from}} \cite{whalen2019}).

As evident from Fig. 5, experimental observations match theoretical predictions well. For a Bose gas and small values of $r/\lambda_{\text{dB}}$, ${\textsl{g}}^{(2)}{(r)}$ approaches 2, in agreement with earlier work \cite{whalen2019}. The predicted  3$!$ increase is observed in ${\textsl{g}}^{(3)}{(r)}$. In contrast, for the Fermi gas, molecule formation is strongly suppressed at small values of $r/\lambda_{\text{dB}}$, and more strongly so for trimers. Correlations decay towards unity on the length scale of $\lambda_{\text{dB}}$ as expected.
\begin{figure}
    \centering
    \twocolumngrid
    \includegraphics[width  = 0.48\textwidth]{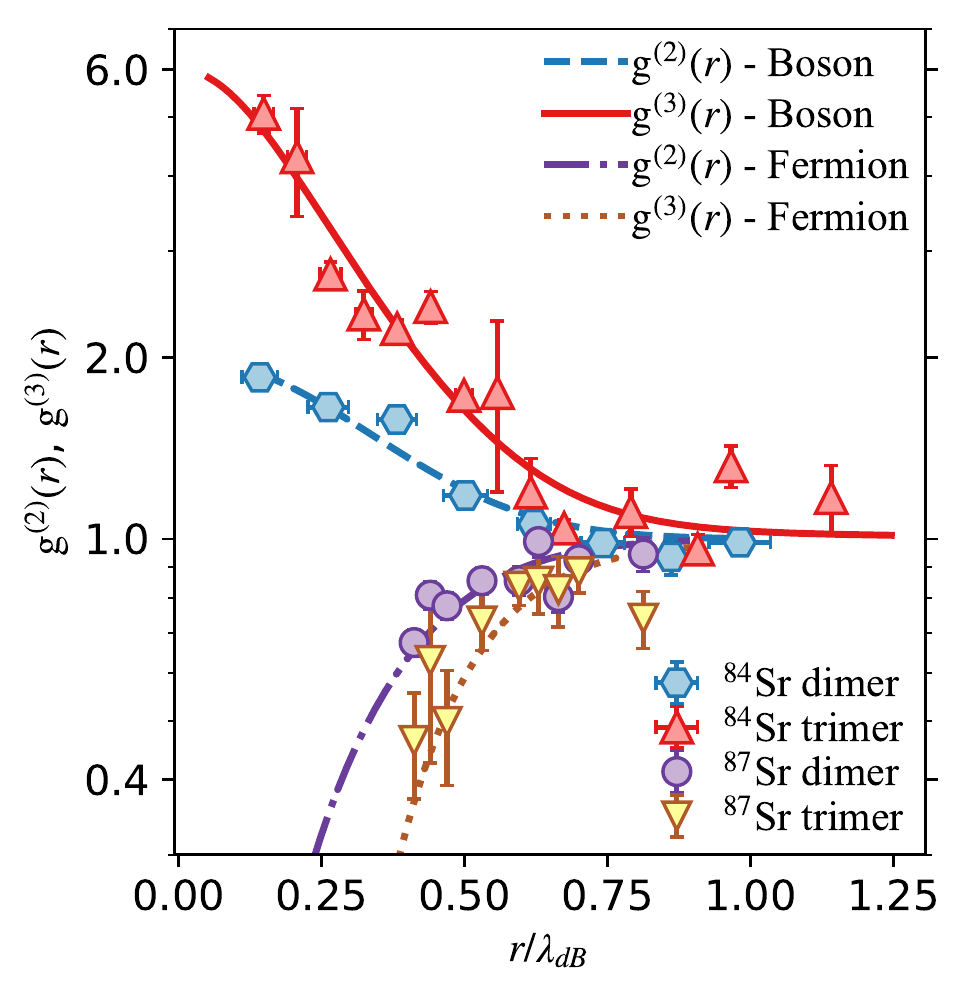}
    \caption{Measured and calculated values of ${\textsl{g}}^{(3)}{(r)}$ and ${\textsl{g}}^{(2)}{(r)}$ for ultracold gases of bosonic ($^{84}\text{Sr}$) and spin-polarized fermionic ($^{87}\text{Sr}$) atoms. Each set of experimental measurements is fit to the corresponding theoretical predictions using a single amplitude scaling factor, and the molecular size is scaled by the atomic thermal de Broglie wavelength. 
    Error bars denote standard error of the mean of multiple measurements or uncertainty in $r/\lambda_{\text{dB}}$ due to uncertainty in sample temperature.
    }
    \label{fig:g2g3AllData}
\end{figure}
\section{CONCLUSIONS}
We have demonstrated that measurements of the formation of ground-state trimer ULRMs provide a sensitive $in\,situ$ probe of three-body, nonlocal spatial correlations in ultracold gases, and have applied this probe to observe  bunching and anti-bunching in thermal gases of indistinguishable bosons and fermions respectively. Even higher-order correlations are accessible by observing formation of tetramers and higher $p$-mers \cite{camargo2018}, offering the possibility of comprehensive characterization of correlations in many-body quantum systems.
It should be possible to apply this technique to systems where interactions affect particle correlations, such as in strongly-interacting 1-D gases.

\section{ACKNOWLEDGEMENTS}
Research supported by the AFOSR under Grant No. FA9550-14-1-0007, the NSF under Grant No. 1600059, and the FWF (Austria) under Grant No. FWF SFB-SFB041 ViCom and the FWF Doctoral College W 1243.


\bibliographystyle{apsrev4-2}

%


\begin{thebibliography}{22}%
\makeatletter
\providecommand \@ifxundefined [1]{%
 \@ifx{#1\undefined}
}%
\providecommand \@ifnum [1]{%
 \ifnum #1\expandafter \@firstoftwo
 \else \expandafter \@secondoftwo
 \fi
}%
\providecommand \@ifx [1]{%
 \ifx #1\expandafter \@firstoftwo
 \else \expandafter \@secondoftwo
 \fi
}%
\providecommand \natexlab [1]{#1}%
\providecommand \enquote  [1]{``#1''}%
\providecommand \bibnamefont  [1]{#1}%
\providecommand \bibfnamefont [1]{#1}%
\providecommand \citenamefont [1]{#1}%
\providecommand \href@noop [0]{\@secondoftwo}%
\providecommand \href [0]{\begingroup \@sanitize@url \@href}%
\providecommand \@href[1]{\@@startlink{#1}\@@href}%
\providecommand \@@href[1]{\endgroup#1\@@endlink}%
\providecommand \@sanitize@url [0]{\catcode `\\12\catcode `\$12\catcode
  `\&12\catcode `\#12\catcode `\^12\catcode `\_12\catcode `\%12\relax}%
\providecommand \@@startlink[1]{}%
\providecommand \@@endlink[0]{}%
\providecommand \url  [0]{\begingroup\@sanitize@url \@url }%
\providecommand \@url [1]{\endgroup\@href {#1}{\urlprefix }}%
\providecommand \urlprefix  [0]{URL }%
\providecommand \Eprint [0]{\href }%
\providecommand \doibase [0]{https://doi.org/}%
\providecommand \selectlanguage [0]{\@gobble}%
\providecommand \bibinfo  [0]{\@secondoftwo}%
\providecommand \bibfield  [0]{\@secondoftwo}%
\providecommand \translation [1]{[#1]}%
\providecommand \BibitemOpen [0]{}%
\providecommand \bibitemStop [0]{}%
\providecommand \bibitemNoStop [0]{.\EOS\space}%
\providecommand \EOS [0]{\spacefactor3000\relax}%
\providecommand \BibitemShut  [1]{\csname bibitem#1\endcsname}%
\let\auto@bib@innerbib\@empty
\bibitem [{\citenamefont {Burt}\ \emph {et~al.}(1997)\citenamefont {Burt},
  \citenamefont {Ghrist}, \citenamefont {Myatt}, \citenamefont {Holland},
  \citenamefont {Cornell},\ and\ \citenamefont {Wieman}}]{burt1997}%
  \BibitemOpen
  \bibfield  {author} {\bibinfo {author} {\bibfnamefont {E.~A.}\ \bibnamefont
  {Burt}}, \bibinfo {author} {\bibfnamefont {R.~W.}\ \bibnamefont {Ghrist}},
  \bibinfo {author} {\bibfnamefont {C.~J.}\ \bibnamefont {Myatt}}, \bibinfo
  {author} {\bibfnamefont {M.~J.}\ \bibnamefont {Holland}}, \bibinfo {author}
  {\bibfnamefont {E.~A.}\ \bibnamefont {Cornell}},\ and\ \bibinfo {author}
  {\bibfnamefont {C.~E.}\ \bibnamefont {Wieman}},\ }\href@noop {} {\bibfield
  {journal} {\bibinfo  {journal} {Phys. Rev. Lett.}\ }\textbf {\bibinfo
  {volume} {79}},\ \bibinfo {pages} {337} (\bibinfo {year} {1997})}\BibitemShut
  {NoStop}%
\bibitem [{\citenamefont {Carcy}\ \emph {et~al.}(2019)\citenamefont {Carcy},
  \citenamefont {Cayla}, \citenamefont {Tenart}, \citenamefont {Aspect},
  \citenamefont {Mancini},\ and\ \citenamefont {Cl{\'e}ment}}]{carcy2019}%
  \BibitemOpen
  \bibfield  {author} {\bibinfo {author} {\bibfnamefont {C.}~\bibnamefont
  {Carcy}}, \bibinfo {author} {\bibfnamefont {H.}~\bibnamefont {Cayla}},
  \bibinfo {author} {\bibfnamefont {A.}~\bibnamefont {Tenart}}, \bibinfo
  {author} {\bibfnamefont {A.}~\bibnamefont {Aspect}}, \bibinfo {author}
  {\bibfnamefont {M.}~\bibnamefont {Mancini}},\ and\ \bibinfo {author}
  {\bibfnamefont {D.}~\bibnamefont {Cl{\'e}ment}},\ }\href@noop {} {\bibfield
  {journal} {\bibinfo  {journal} {Phys. Rev. X}\ }\textbf {\bibinfo {volume}
  {9}},\ \bibinfo {pages} {041028} (\bibinfo {year} {2019})}\BibitemShut
  {NoStop}%
\bibitem [{\citenamefont {Browaeys}\ and\ \citenamefont
  {Lahaye}(2020)}]{browaeys2020}%
  \BibitemOpen
  \bibfield  {author} {\bibinfo {author} {\bibfnamefont {A.}~\bibnamefont
  {Browaeys}}\ and\ \bibinfo {author} {\bibfnamefont {T.}~\bibnamefont
  {Lahaye}},\ }\href@noop {} {\bibfield  {journal} {\bibinfo  {journal} {Nat.
  Phys.}\ }\textbf {\bibinfo {volume} {16}},\ \bibinfo {pages} {132} (\bibinfo
  {year} {2020})}\BibitemShut {NoStop}%
\bibitem [{\citenamefont {Semeghini}\ \emph {et~al.}(2021)\citenamefont
  {Semeghini}, \citenamefont {Levine}, \citenamefont {Keesling}, \citenamefont
  {Ebadi}, \citenamefont {Wang}, \citenamefont {Bluvstein}, \citenamefont
  {Verresen}, \citenamefont {Pichler}, \citenamefont {Kalinowski},
  \citenamefont {Samajdar}, \citenamefont {Omran}, \citenamefont {Sachdev},
  \citenamefont {Vishwanath}, \citenamefont {Greiner}, \citenamefont
  {Vuleti{\'c}},\ and\ \citenamefont {Lukin}}]{semeghini2021}%
  \BibitemOpen
  \bibfield  {author} {\bibinfo {author} {\bibfnamefont {G.}~\bibnamefont
  {Semeghini}}, \bibinfo {author} {\bibfnamefont {H.}~\bibnamefont {Levine}},
  \bibinfo {author} {\bibfnamefont {A.}~\bibnamefont {Keesling}}, \bibinfo
  {author} {\bibfnamefont {S.}~\bibnamefont {Ebadi}}, \bibinfo {author}
  {\bibfnamefont {T.~T.}\ \bibnamefont {Wang}}, \bibinfo {author}
  {\bibfnamefont {D.}~\bibnamefont {Bluvstein}}, \bibinfo {author}
  {\bibfnamefont {R.}~\bibnamefont {Verresen}}, \bibinfo {author}
  {\bibfnamefont {H.}~\bibnamefont {Pichler}}, \bibinfo {author} {\bibfnamefont
  {M.}~\bibnamefont {Kalinowski}}, \bibinfo {author} {\bibfnamefont
  {R.}~\bibnamefont {Samajdar}}, \bibinfo {author} {\bibfnamefont
  {A.}~\bibnamefont {Omran}}, \bibinfo {author} {\bibfnamefont
  {S.}~\bibnamefont {Sachdev}}, \bibinfo {author} {\bibfnamefont
  {A.}~\bibnamefont {Vishwanath}}, \bibinfo {author} {\bibfnamefont
  {M.}~\bibnamefont {Greiner}}, \bibinfo {author} {\bibfnamefont
  {V.}~\bibnamefont {Vuleti{\'c}}},\ and\ \bibinfo {author} {\bibfnamefont
  {M.~D.}\ \bibnamefont {Lukin}},\ }\href@noop {} {\bibfield  {journal}
  {\bibinfo  {journal} {Science}\ }\textbf {\bibinfo {volume} {374}},\ \bibinfo
  {pages} {1242} (\bibinfo {year} {2021})}\BibitemShut {NoStop}%
\bibitem [{\citenamefont {Mazurenko}\ \emph {et~al.}(2017)\citenamefont
  {Mazurenko}, \citenamefont {Chiu}, \citenamefont {Ji}, \citenamefont
  {Parsons}, \citenamefont {{Kan{\'a}sz-Nagy}}, \citenamefont {Schmidt},
  \citenamefont {Grusdt}, \citenamefont {Demler}, \citenamefont {Greif},\ and\
  \citenamefont {Greiner}}]{mazurenko2017}%
  \BibitemOpen
  \bibfield  {author} {\bibinfo {author} {\bibfnamefont {A.}~\bibnamefont
  {Mazurenko}}, \bibinfo {author} {\bibfnamefont {C.~S.}\ \bibnamefont {Chiu}},
  \bibinfo {author} {\bibfnamefont {G.}~\bibnamefont {Ji}}, \bibinfo {author}
  {\bibfnamefont {M.~F.}\ \bibnamefont {Parsons}}, \bibinfo {author}
  {\bibfnamefont {M.}~\bibnamefont {{Kan{\'a}sz-Nagy}}}, \bibinfo {author}
  {\bibfnamefont {R.}~\bibnamefont {Schmidt}}, \bibinfo {author} {\bibfnamefont
  {F.}~\bibnamefont {Grusdt}}, \bibinfo {author} {\bibfnamefont
  {E.}~\bibnamefont {Demler}}, \bibinfo {author} {\bibfnamefont
  {D.}~\bibnamefont {Greif}},\ and\ \bibinfo {author} {\bibfnamefont
  {M.}~\bibnamefont {Greiner}},\ }\href@noop {} {\bibfield  {journal} {\bibinfo
   {journal} {Nature}\ }\textbf {\bibinfo {volume} {545}},\ \bibinfo {pages}
  {462} (\bibinfo {year} {2017})}\BibitemShut {NoStop}%
\bibitem [{\citenamefont {Hart}\ \emph {et~al.}(2015)\citenamefont {Hart},
  \citenamefont {Duarte}, \citenamefont {Yang}, \citenamefont {Liu},
  \citenamefont {Paiva}, \citenamefont {Khatami}, \citenamefont {Scalettar},
  \citenamefont {Trivedi}, \citenamefont {Huse},\ and\ \citenamefont
  {Hulet}}]{hart2015}%
  \BibitemOpen
  \bibfield  {author} {\bibinfo {author} {\bibfnamefont {R.~A.}\ \bibnamefont
  {Hart}}, \bibinfo {author} {\bibfnamefont {P.~M.}\ \bibnamefont {Duarte}},
  \bibinfo {author} {\bibfnamefont {T.-L.}\ \bibnamefont {Yang}}, \bibinfo
  {author} {\bibfnamefont {X.}~\bibnamefont {Liu}}, \bibinfo {author}
  {\bibfnamefont {T.}~\bibnamefont {Paiva}}, \bibinfo {author} {\bibfnamefont
  {E.}~\bibnamefont {Khatami}}, \bibinfo {author} {\bibfnamefont {R.~T.}\
  \bibnamefont {Scalettar}}, \bibinfo {author} {\bibfnamefont {N.}~\bibnamefont
  {Trivedi}}, \bibinfo {author} {\bibfnamefont {D.~A.}\ \bibnamefont {Huse}},\
  and\ \bibinfo {author} {\bibfnamefont {R.~G.}\ \bibnamefont {Hulet}},\
  }\href@noop {} {\bibfield  {journal} {\bibinfo  {journal} {Nature}\ }\textbf
  {\bibinfo {volume} {519}},\ \bibinfo {pages} {211} (\bibinfo {year}
  {2015})}\BibitemShut {NoStop}%
\bibitem [{\citenamefont {Fletcher}\ \emph {et~al.}(2017)\citenamefont
  {Fletcher}, \citenamefont {Lopes}, \citenamefont {Man}, \citenamefont
  {Navon}, \citenamefont {Smith}, \citenamefont {Zwierlein},\ and\
  \citenamefont {Hadzibabic}}]{fletcher2017}%
  \BibitemOpen
  \bibfield  {author} {\bibinfo {author} {\bibfnamefont {R.~J.}\ \bibnamefont
  {Fletcher}}, \bibinfo {author} {\bibfnamefont {R.}~\bibnamefont {Lopes}},
  \bibinfo {author} {\bibfnamefont {J.}~\bibnamefont {Man}}, \bibinfo {author}
  {\bibfnamefont {N.}~\bibnamefont {Navon}}, \bibinfo {author} {\bibfnamefont
  {R.~P.}\ \bibnamefont {Smith}}, \bibinfo {author} {\bibfnamefont {M.~W.}\
  \bibnamefont {Zwierlein}},\ and\ \bibinfo {author} {\bibfnamefont
  {Z.}~\bibnamefont {Hadzibabic}},\ }\href@noop {} {\bibfield  {journal}
  {\bibinfo  {journal} {Science}\ }\textbf {\bibinfo {volume} {355}},\ \bibinfo
  {pages} {377} (\bibinfo {year} {2017})}\BibitemShut {NoStop}%
\bibitem [{\citenamefont {Tolra}\ \emph {et~al.}(2004)\citenamefont {Tolra},
  \citenamefont {O'Hara}, \citenamefont {Huckans}, \citenamefont {Phillips},
  \citenamefont {Rolston},\ and\ \citenamefont {Porto}}]{tolra2004}%
  \BibitemOpen
  \bibfield  {author} {\bibinfo {author} {\bibfnamefont {B.~L.}\ \bibnamefont
  {Tolra}}, \bibinfo {author} {\bibfnamefont {K.~M.}\ \bibnamefont {O'Hara}},
  \bibinfo {author} {\bibfnamefont {J.~H.}\ \bibnamefont {Huckans}}, \bibinfo
  {author} {\bibfnamefont {W.~D.}\ \bibnamefont {Phillips}}, \bibinfo {author}
  {\bibfnamefont {S.~L.}\ \bibnamefont {Rolston}},\ and\ \bibinfo {author}
  {\bibfnamefont {J.~V.}\ \bibnamefont {Porto}},\ }\href@noop {} {\bibfield
  {journal} {\bibinfo  {journal} {Phys. Rev. Lett.}\ }\textbf {\bibinfo
  {volume} {92}},\ \bibinfo {pages} {190401} (\bibinfo {year}
  {2004})}\BibitemShut {NoStop}%
\bibitem [{\citenamefont {Kinoshita}\ \emph {et~al.}(2005)\citenamefont
  {Kinoshita}, \citenamefont {Wenger},\ and\ \citenamefont
  {Weiss}}]{kinoshita2005}%
  \BibitemOpen
  \bibfield  {author} {\bibinfo {author} {\bibfnamefont {T.}~\bibnamefont
  {Kinoshita}}, \bibinfo {author} {\bibfnamefont {T.}~\bibnamefont {Wenger}},\
  and\ \bibinfo {author} {\bibfnamefont {D.~S.}\ \bibnamefont {Weiss}},\
  }\href@noop {} {\bibfield  {journal} {\bibinfo  {journal} {Phys. Rev. Lett.}\
  }\textbf {\bibinfo {volume} {95}},\ \bibinfo {pages} {190406} (\bibinfo
  {year} {2005})}\BibitemShut {NoStop}%
\bibitem [{\citenamefont {Haller}\ \emph {et~al.}(2011)\citenamefont {Haller},
  \citenamefont {Rabie}, \citenamefont {Mark}, \citenamefont {Danzl},
  \citenamefont {Hart}, \citenamefont {Lauber}, \citenamefont {Pupillo},\ and\
  \citenamefont {N{\"a}gerl}}]{haller2011}%
  \BibitemOpen
  \bibfield  {author} {\bibinfo {author} {\bibfnamefont {E.}~\bibnamefont
  {Haller}}, \bibinfo {author} {\bibfnamefont {M.}~\bibnamefont {Rabie}},
  \bibinfo {author} {\bibfnamefont {M.~J.}\ \bibnamefont {Mark}}, \bibinfo
  {author} {\bibfnamefont {J.~G.}\ \bibnamefont {Danzl}}, \bibinfo {author}
  {\bibfnamefont {R.}~\bibnamefont {Hart}}, \bibinfo {author} {\bibfnamefont
  {K.}~\bibnamefont {Lauber}}, \bibinfo {author} {\bibfnamefont
  {G.}~\bibnamefont {Pupillo}},\ and\ \bibinfo {author} {\bibfnamefont {H.-C.}\
  \bibnamefont {N{\"a}gerl}},\ }\href@noop {} {\bibfield  {journal} {\bibinfo
  {journal} {Phys. Rev. Lett.}\ }\textbf {\bibinfo {volume} {107}},\ \bibinfo
  {pages} {230404} (\bibinfo {year} {2011})}\BibitemShut {NoStop}%
\bibitem [{\citenamefont {Bakr}\ \emph {et~al.}(2009)\citenamefont {Bakr},
  \citenamefont {Gillen}, \citenamefont {Peng}, \citenamefont {F{\"o}lling},\
  and\ \citenamefont {Greiner}}]{bakr2009}%
  \BibitemOpen
  \bibfield  {author} {\bibinfo {author} {\bibfnamefont {W.~S.}\ \bibnamefont
  {Bakr}}, \bibinfo {author} {\bibfnamefont {J.~I.}\ \bibnamefont {Gillen}},
  \bibinfo {author} {\bibfnamefont {A.}~\bibnamefont {Peng}}, \bibinfo {author}
  {\bibfnamefont {S.}~\bibnamefont {F{\"o}lling}},\ and\ \bibinfo {author}
  {\bibfnamefont {M.}~\bibnamefont {Greiner}},\ }\href@noop {} {\bibfield
  {journal} {\bibinfo  {journal} {Nature}\ }\textbf {\bibinfo {volume} {462}},\
  \bibinfo {pages} {74} (\bibinfo {year} {2009})}\BibitemShut {NoStop}%
\bibitem [{\citenamefont {Whalen}\ \emph {et~al.}(2019)\citenamefont {Whalen},
  \citenamefont {Kanungo}, \citenamefont {Ding}, \citenamefont {Wagner},
  \citenamefont {Schmidt}, \citenamefont {Sadeghpour}, \citenamefont {Yoshida},
  \citenamefont {Burgd{\"o}rfer}, \citenamefont {Dunning},\ and\ \citenamefont
  {Killian}}]{whalen2019}%
  \BibitemOpen
  \bibfield  {author} {\bibinfo {author} {\bibfnamefont {J.~D.}\ \bibnamefont
  {Whalen}}, \bibinfo {author} {\bibfnamefont {S.~K.}\ \bibnamefont {Kanungo}},
  \bibinfo {author} {\bibfnamefont {R.}~\bibnamefont {Ding}}, \bibinfo {author}
  {\bibfnamefont {M.}~\bibnamefont {Wagner}}, \bibinfo {author} {\bibfnamefont
  {R.}~\bibnamefont {Schmidt}}, \bibinfo {author} {\bibfnamefont {H.~R.}\
  \bibnamefont {Sadeghpour}}, \bibinfo {author} {\bibfnamefont
  {S.}~\bibnamefont {Yoshida}}, \bibinfo {author} {\bibfnamefont
  {J.}~\bibnamefont {Burgd{\"o}rfer}}, \bibinfo {author} {\bibfnamefont
  {F.~B.}\ \bibnamefont {Dunning}},\ and\ \bibinfo {author} {\bibfnamefont
  {T.~C.}\ \bibnamefont {Killian}},\ }\href@noop {} {\bibfield  {journal}
  {\bibinfo  {journal} {Phys. Rev. A}\ }\textbf {\bibinfo {volume} {100}},\
  \bibinfo {pages} {011402} (\bibinfo {year} {2019})}\BibitemShut {NoStop}%
\bibitem [{\citenamefont {Greene}\ \emph {et~al.}(2000)\citenamefont {Greene},
  \citenamefont {Dickinson},\ and\ \citenamefont {Sadeghpour}}]{greene2000}%
  \BibitemOpen
  \bibfield  {author} {\bibinfo {author} {\bibfnamefont {C.~H.}\ \bibnamefont
  {Greene}}, \bibinfo {author} {\bibfnamefont {A.~S.}\ \bibnamefont
  {Dickinson}},\ and\ \bibinfo {author} {\bibfnamefont {H.~R.}\ \bibnamefont
  {Sadeghpour}},\ }\href@noop {} {\bibfield  {journal} {\bibinfo  {journal}
  {Phys. Rev. Lett.}\ }\textbf {\bibinfo {volume} {85}},\ \bibinfo {pages}
  {2458} (\bibinfo {year} {2000})}\BibitemShut {NoStop}%
\bibitem [{\citenamefont {Bendkowsky}\ \emph {et~al.}(2009)\citenamefont
  {Bendkowsky}, \citenamefont {Butscher}, \citenamefont {Nipper}, \citenamefont
  {Shaffer}, \citenamefont {L{\"o}w},\ and\ \citenamefont
  {Pfau}}]{bendkowsky2009}%
  \BibitemOpen
  \bibfield  {author} {\bibinfo {author} {\bibfnamefont {V.}~\bibnamefont
  {Bendkowsky}}, \bibinfo {author} {\bibfnamefont {B.}~\bibnamefont
  {Butscher}}, \bibinfo {author} {\bibfnamefont {J.}~\bibnamefont {Nipper}},
  \bibinfo {author} {\bibfnamefont {J.~P.}\ \bibnamefont {Shaffer}}, \bibinfo
  {author} {\bibfnamefont {R.}~\bibnamefont {L{\"o}w}},\ and\ \bibinfo {author}
  {\bibfnamefont {T.}~\bibnamefont {Pfau}},\ }\href@noop {} {\bibfield
  {journal} {\bibinfo  {journal} {Nature}\ }\textbf {\bibinfo {volume} {458}},\
  \bibinfo {pages} {1005} (\bibinfo {year} {2009})}\BibitemShut {NoStop}%
\bibitem [{\citenamefont {Stellmer}\ \emph {et~al.}(2013)\citenamefont
  {Stellmer}, \citenamefont {Schreck},\ and\ \citenamefont
  {Killian}}]{stellmer2013}%
  \BibitemOpen
  \bibfield  {author} {\bibinfo {author} {\bibfnamefont {S.}~\bibnamefont
  {Stellmer}}, \bibinfo {author} {\bibfnamefont {F.}~\bibnamefont {Schreck}},\
  and\ \bibinfo {author} {\bibfnamefont {T.~C.}\ \bibnamefont {Killian}},\ }in\
  \href@noop {} {\emph {\bibinfo {booktitle} {Annual {{Review}} of {{Cold
  Atoms}} and {{Molecules}}}}},\ \bibinfo {series} {Annual {{Review}} of {{Cold
  Atoms}} and {{Molecules}}}, Vol.\ \bibinfo {volume} {Volume 2}\ (\bibinfo
  {year} {2013})\ pp.\ \bibinfo {pages} {1--80}\BibitemShut {NoStop}%
\bibitem [{\citenamefont {{de Escobar}}\ \emph {et~al.}(2009)\citenamefont {{de
  Escobar}}, \citenamefont {Mickelson}, \citenamefont {Yan}, \citenamefont
  {DeSalvo}, \citenamefont {Nagel},\ and\ \citenamefont
  {Killian}}]{deescobar2009}%
  \BibitemOpen
  \bibfield  {author} {\bibinfo {author} {\bibfnamefont {Y.~N.~M.}\
  \bibnamefont {{de Escobar}}}, \bibinfo {author} {\bibfnamefont {P.~G.}\
  \bibnamefont {Mickelson}}, \bibinfo {author} {\bibfnamefont {M.}~\bibnamefont
  {Yan}}, \bibinfo {author} {\bibfnamefont {B.~J.}\ \bibnamefont {DeSalvo}},
  \bibinfo {author} {\bibfnamefont {S.~B.}\ \bibnamefont {Nagel}},\ and\
  \bibinfo {author} {\bibfnamefont {T.~C.}\ \bibnamefont {Killian}},\
  }\href@noop {} {\bibfield  {journal} {\bibinfo  {journal} {Phys. Rev. Lett.}\
  }\textbf {\bibinfo {volume} {103}},\ \bibinfo {pages} {200402} (\bibinfo
  {year} {2009})}\BibitemShut {NoStop}%
\bibitem [{\citenamefont {O'Hara}\ \emph {et~al.}(2001)\citenamefont {O'Hara},
  \citenamefont {Gehm}, \citenamefont {Granade},\ and\ \citenamefont
  {Thomas}}]{ohara2001}%
  \BibitemOpen
  \bibfield  {author} {\bibinfo {author} {\bibfnamefont {K.~M.}\ \bibnamefont
  {O'Hara}}, \bibinfo {author} {\bibfnamefont {M.~E.}\ \bibnamefont {Gehm}},
  \bibinfo {author} {\bibfnamefont {S.~R.}\ \bibnamefont {Granade}},\ and\
  \bibinfo {author} {\bibfnamefont {J.~E.}\ \bibnamefont {Thomas}},\
  }\href@noop {} {\bibfield  {journal} {\bibinfo  {journal} {Phys. Rev. A}\
  }\textbf {\bibinfo {volume} {64}},\ \bibinfo {pages} {051403} (\bibinfo
  {year} {2001})}\BibitemShut {NoStop}%
\bibitem [{\citenamefont {Gallagher}(1994)}]{gallagher1994}%
  \BibitemOpen
  \bibfield  {author} {\bibinfo {author} {\bibfnamefont {T.~F.}\ \bibnamefont
  {Gallagher}},\ }\href@noop {} {}Cambridge {{Monographs}} on {{Atomic}},
  {{Molecular}} and {{Chemical Physics}}\ (\bibinfo {address} {{Cambridge}},\
  \bibinfo {year} {1994})\BibitemShut {NoStop}%
\bibitem [{\citenamefont {Naraschewski}\ and\ \citenamefont
  {Glauber}(1999)}]{naraschewski1999}%
  \BibitemOpen
  \bibfield  {author} {\bibinfo {author} {\bibfnamefont {M.}~\bibnamefont
  {Naraschewski}}\ and\ \bibinfo {author} {\bibfnamefont {R.~J.}\ \bibnamefont
  {Glauber}},\ }\href@noop {} {\bibfield  {journal} {\bibinfo  {journal} {Phys.
  Rev. A}\ }\textbf {\bibinfo {volume} {59}},\ \bibinfo {pages} {4595}
  (\bibinfo {year} {1999})}\BibitemShut {NoStop}%
\bibitem [{\citenamefont {Wick}(1950)}]{wick1950}%
  \BibitemOpen
  \bibfield  {author} {\bibinfo {author} {\bibfnamefont {G.~C.}\ \bibnamefont
  {Wick}},\ }\href@noop {} {\bibfield  {journal} {\bibinfo  {journal} {Phys.
  Rev.}\ }\textbf {\bibinfo {volume} {80}},\ \bibinfo {pages} {268} (\bibinfo
  {year} {1950})}\BibitemShut {NoStop}%
\bibitem [{\citenamefont {Hodgman}\ \emph {et~al.}(2011)\citenamefont
  {Hodgman}, \citenamefont {Dall}, \citenamefont {Manning}, \citenamefont
  {Baldwin},\ and\ \citenamefont {Truscott}}]{hodgman2011}%
  \BibitemOpen
  \bibfield  {author} {\bibinfo {author} {\bibfnamefont {S.~S.}\ \bibnamefont
  {Hodgman}}, \bibinfo {author} {\bibfnamefont {R.~G.}\ \bibnamefont {Dall}},
  \bibinfo {author} {\bibfnamefont {A.~G.}\ \bibnamefont {Manning}}, \bibinfo
  {author} {\bibfnamefont {K.~G.~H.}\ \bibnamefont {Baldwin}},\ and\ \bibinfo
  {author} {\bibfnamefont {A.~G.}\ \bibnamefont {Truscott}},\ }\href@noop {}
  {\bibfield  {journal} {\bibinfo  {journal} {Science}\ }\textbf {\bibinfo
  {volume} {331}},\ \bibinfo {pages} {1046} (\bibinfo {year}
  {2011})}\BibitemShut {NoStop}%
\bibitem [{\citenamefont {Camargo}\ \emph {et~al.}(2018)\citenamefont
  {Camargo}, \citenamefont {Schmidt}, \citenamefont {Whalen}, \citenamefont
  {Ding}, \citenamefont {Woehl}, \citenamefont {Yoshida}, \citenamefont
  {Burgd{\"o}rfer}, \citenamefont {Dunning}, \citenamefont {Sadeghpour},
  \citenamefont {Demler},\ and\ \citenamefont {Killian}}]{camargo2018}%
  \BibitemOpen
  \bibfield  {author} {\bibinfo {author} {\bibfnamefont {F.}~\bibnamefont
  {Camargo}}, \bibinfo {author} {\bibfnamefont {R.}~\bibnamefont {Schmidt}},
  \bibinfo {author} {\bibfnamefont {J.~D.}\ \bibnamefont {Whalen}}, \bibinfo
  {author} {\bibfnamefont {R.}~\bibnamefont {Ding}}, \bibinfo {author}
  {\bibfnamefont {G.}~\bibnamefont {Woehl}}, \bibinfo {author} {\bibfnamefont
  {S.}~\bibnamefont {Yoshida}}, \bibinfo {author} {\bibfnamefont
  {J.}~\bibnamefont {Burgd{\"o}rfer}}, \bibinfo {author} {\bibfnamefont
  {F.~B.}\ \bibnamefont {Dunning}}, \bibinfo {author} {\bibfnamefont {H.~R.}\
  \bibnamefont {Sadeghpour}}, \bibinfo {author} {\bibfnamefont
  {E.}~\bibnamefont {Demler}},\ and\ \bibinfo {author} {\bibfnamefont {T.~C.}\
  \bibnamefont {Killian}},\ }\href@noop {} {\bibfield  {journal} {\bibinfo
  {journal} {Phys. Rev. Lett.}\ }\textbf {\bibinfo {volume} {120}},\ \bibinfo
  {pages} {083401} (\bibinfo {year} {2018})}\BibitemShut {NoStop}%
\end{thebibliography}
\end{document}